\documentclass[aps,prl,twocolumn,groupedaddress]{revtex4}
\usepackage{graphicx}
\usepackage{epstopdf}

\begin{document}


\title{Can amphiphile architecture directly control vesicle size?}

\author{Martin J.~Greenall}
\affiliation{Institut Charles Sadron, University of Strasbourg, CNRS -
UPR 22, 23, rue du Loess, 67034 Strasbourg, France}
\author{Carlos M.~Marques}
\affiliation{Institut Charles Sadron, University of Strasbourg, CNRS -
UPR 22, 23, rue du Loess, 67034 Strasbourg, France}

\date{\today}

\begin{abstract}
Bilayer membranes self-assembled from simple amphiphiles in solution always have a planar
ground-state shape. This is a consequence of several
internal relaxation mechanisms of the membrane and prevents the
straightforward control of vesicle size. Here, we show that this principle can be circumvented and
that direct size control by molecular design is a realistic
possibility. Using coarse-grained calculations, we design tetrablock
copolymers that form membranes with a
preferred curvature, and demonstrate how to form low-polydispersity vesicles while suppressing micellization.
\end{abstract}

\pacs{05.20.-y, 05.70.-a, 82.35.Jk, 82.70.Uv}

\maketitle

A fundamental process in soft matter science is the self-assembly of
amphiphilic molecules into structures ranging from simple micelles to complex connected aggregates
\cite{he_schmid,jain_bates,zhang_eisenberg}. Self-assembled structures not only occur naturally
in living cells, but can also be designed for applications such
as drug delivery \cite{haleva_diamant}. The question that lies at the heart of this field is how the properties of the individual amphiphilic
molecules control the topology of the aggregates they form \cite{choi}. One of the
major unsolved problems is to design a molecule that
can directly fix the curvature of a membrane in solution. In addition
to its fundamental interest, this question is of great practical
importance, as finding such a molecule would allow the spontaneous formation of vesicles of a well-defined size, yielding
precise control of drug delivery systems.

At present, membrane curvature can only be controlled by rather complex
procedures. Several of these
\cite{lee_agarwal,kaler,katagiri,li_cohen-stuart,nieh,joannic}
blend two species of amphiphile \cite{campelo}, so that the symmetry
of the inner and
outer bilayer leaflets is broken
\cite{safran_pincus_andelman} and the vesicle has a preferred radius. Such methods have the disadvantage that
blends of amphiphiles can form a wide range of micelles, which may coexist with the target vesicle structure
\cite{li_cohen-stuart}. Other methods involve the use of more intricate
vesicle formation pathways, such as dewetting from a template
\cite{howse}, cooling and
warming through a cylinder-vesicle shape transition \cite{rank},
electroformation on micropatterned glass slides \cite{taylor}, flow
focusing \cite{thiele}, and combined extrusion and
dialysis \cite{zhu}.

\begin{figure}
\includegraphics[width=2.9in]{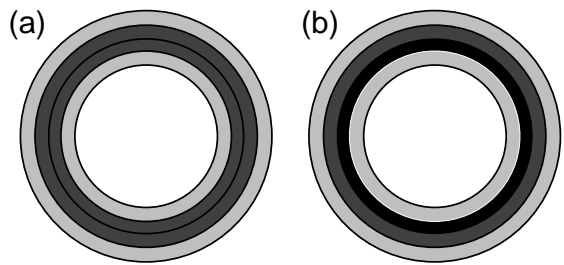}
\caption{\label{ditetra_fig} (a) Equatorial section of a bilayer vesicle of diblock
  copolymers. Hydrophilic blocks are light gray and hydrophobic blocks
  are dark gray. (b) Monolayer vesicle of tetrablock
copolymers. The hydrophobic blocks are colored black and
dark gray, with black block being more hydrophobic.}
\end{figure}

In this Letter, we investigate an alternative strategy for controlling
membrane curvature. We break the membrane symmetry by the use of ABCA$^\prime$ tetrablock
copolymers \cite{brannan_bates,gomez,cui}. The outer A and A$^\prime$ blocks of the
polymer are formed of the same hydrophilic material, and the B and C
blocks are hydrophobic and
have a repulsive interaction with each other. In contrast to the bilayers formed by
diblocks (Fig.\ \ref{ditetra_fig}a), these molecules form
asymmetric {\em monolayers} in solution \cite{brannan_bates} (Fig.\ \ref{ditetra_fig}b). We use tetrablocks rather
than ABC triblocks since, in this latter case, the A and C blocks
would have to be hydrophilic and have a strong mutual repulsion for
asymmetric monolayers to form. This
combination is hard to achieve, both because of the difficulty of
finding hydrophilic compounds that repel strongly and the dilution of the hydrophilic layers by solvent,
which weakens any interaction between them.

Tetrablock vesicles have indeed been formed in preliminary experimental
investigations by Brannan and Bates \cite{brannan_bates}, who also
achieved some size control of the aggregates. Such vesicles have also
been found in very recent Monte Carlo simulations \cite{cui},
although this method can only access small vesicles and does not allow
size control to be studied. Here, we present
a concrete theoretical demonstration of the basic principle that
tetrablock copolymers can form bilayers of a preferred curvature, and
show how to design these molecules to control the vesicle
radius and polydispersity while avoiding micelle formation \cite{gomez}.

We focus on a simple model of ABCA$^\prime$ tetrablocks in A
homopolymer `solvent'. Dilute block copolymer-homopolymer blends provide a
good model of aqueous copolymer solutions, and show the
same sequence of morphologies as a function of block lengths \cite{kinning_winey_thomas}, since the mechanism
that drives the shape transitions (the chain crowding in the different
layers of the membrane) is the same in both cases. These systems are
well-described by SCFT \cite {gbm_jcp}. Furthermore, simple
mean-field models of copolymer-homopolymer systems
have provided important qualitative insights even
into aqueous solutions of small biological molecules, notably the
problem of membrane fusion \cite{katsov1,papanicolaou}.

To begin, we consider copolymers where all four
segments contain the same number of monomers $N/4$. For simplicity,
the A homopolymer molecules also contain $N/4$
monomers. The strengths of the
interactions between the species are set by Flory $\chi$
parameters. Once two $\chi$ parameters
have been chosen, the third must be calculated from a
relation involving the polarizabilities of the species
\cite{boudenne_book}. To calculate the density profiles and free energies of the
self-assembled structures, we used a simple coarse-grained mean-field theory
(self-consistent field theory, or SCFT). The individual polymer
molecules interact via a contact potential, and are modeled by random
walks, which are averaged over by the SCFT to calculate the density
profiles \cite{edwards,matsen_book}. SCFT is well-adapted to
our current investigation, as its speed allows us to study a much larger range of vesicle
sizes than Monte Carlo methods applied to a comparable
system \cite{cui}, and can be nearly as accurate as these more expensive methods for long polymers \cite{cavallo}.
The diffusion equations describing the
polymers were solved by a
finite-difference method and the SCFT equations by an iterative scheme
\cite{kim_matsen}, supplemented by extrapolation.

Since we focus on spherical vesicles, we
perform many of our calculations assuming spherical symmetry in a
spherical box. We must connect the free energy
of the subsystem of volume $V$ containing the vesicle to that of the
whole system. To do this, we
calculate the free-energy
density $\tilde{F}$ of a box (with periodic boundary conditions) containing a single spherical vesicle in the canonical
ensemble. We then vary the volume of the
simulation box at constant overall
copolymer volume fraction $\phi$ \cite{gbm_jcp}. This
corresponds to a larger system (of fixed total
volume and fixed copolymer volume fraction) varying the number of aggregates and hence the
volume occupied by each. If $\tilde{F}(V)$ has a minimum, this means that the vesicle membranes have an optimum
curvature.

This minimum corresponds to the absolute free energy
minimum of a solution of spherical vesicles, and a point on $\tilde{F}(V)$ corresponds to a {\em monodisperse} solution of vesicles of a
given size. We now use these curves to
take into account simple fluctuations around the free energy minimum
to 
calculate the polydispersity $\Delta$ of the vesicles. This is related to the free energy
$f_p$ of an aggregate of $p$ molecules by
$1/\Delta^2=\partial ^2f_p/\partial p^2$ \cite{puvvada}. To extract this quantity, we write
\begin{equation}
\tilde{F}=(\phi-v_{\text{m}}/V)\ln
  [(\phi-v_{\text{m}}/V)/e]
+(\phi-v_{\text{m}}/V)f_1+vf_p/V
\label{FE_system}
\end{equation}
where $1/V$ is the number density of aggregates and $v_{\text{m}}=pv$ is
the volume of an aggregate. The first term arises
from the entropy of the free copolymers in solution. Now, a single SCFT
calculation finds the local free energy minimum $\tilde{F}(V)$ for a vesicle in a box
of volume
$V$. In the process, it determines the optimum number of
molecules in the vesicle for this box size and so corresponds to
minimizing $\tilde{F}$ with respect to $p$ at a given $1/V$. Varying
$V$ then yields $\tilde{F}(V)$, from which we can
read off $\partial^2\tilde{F}/\partial V^2$. Remembering that this derivative
is evaluated along the line where $\partial\tilde{F}/\partial p|_V=0$,
we find that
\begin{equation}
\frac{1}{\Delta^2}=\frac{\partial^2 f_p}{\partial
  p^2}=\frac{v}{v_{\text{m}}^2/(V^3\partial^2\tilde{F}/\partial
  V^2)-(\phi V-v_{\text{m}})}
\label{polydisp}
\end{equation}
allowing us to calculate $\Delta$.

\begin{figure}
\includegraphics[width=2.9in]{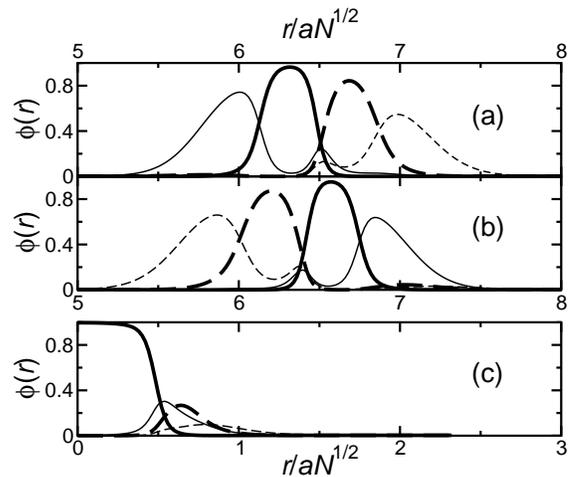}
\caption{\label{profiles_fig} Sample volume fraction profiles for (a)
  ABCA$^\prime$ vesicles, (b) vesicles in the
  A$^\prime$CBA orientation, and (c) micelles. A-blocks: thin full lines,
  B-blocks: thick full lines, C-blocks: thick dashed lines, A$^\prime$-
  blocks: thin dashed lines.}
\end{figure}

We now demonstrate that the target vesicle structure
is a solution to SCFT. In Fig.\ \ref{profiles_fig}a, we plot cuts
through the density profile
of an ABCA$^\prime$ vesicle, with $\phi=0.05$. The $\chi$ parameters must be large
enough for the amphiphile to aggregate, and so we set
$\chi_{\text{AB}}=50/N$ and $\chi_{\text{AC}}=30/N$, where $N$ is the
total number of monomers in the copolymer. So that the B and C species
demix, we choose the larger of the two possible values
\cite{boudenne_book} for $\chi_{\text{BC}}$, which we set to
$157.5/N$. We measure all lengths in units of $aN^{1/2}$, where $a$
is the segment length \cite{matsen_book}.

The ABCA$^\prime$ structure sketched in Fig.\
\ref{ditetra_fig} is clearly reproduced in Fig.\ \ref{profiles_fig}a. The
strongly hydrophobic B-blocks lie in the inner half of the membrane,
so that the more energetically unfavorable AB interface has a smaller
area. Surprisingly, we also find solutions,
shown in Fig.\ \ref{profiles_fig}b,
where the
B-blocks lie on the outside of the membrane.

In Fig.\ \ref{profiles_fig}c, we plot
the density profile of a micelle, formed in a smaller calculation
box. The core is formed from the strongly hydrophobic B
blocks. This structure is
most likely formed as in ABA triblock
micelles \cite{ulrich}, with the copolymers forming a hairpin shape. 

\begin{figure}
\includegraphics[width=2.9in]{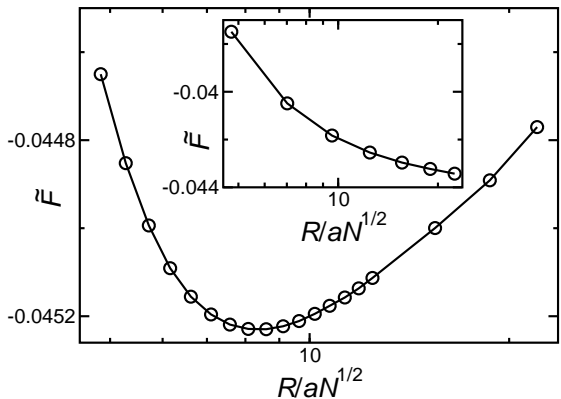}
\caption{\label{min_fig} Free energy density as a
  function of vesicle radius $R$. The inset shows the corresponding
  curve for vesicles formed of polymers in the
  A$^\prime$CBA orientation.}
\end{figure}

In Fig.\ \ref{min_fig}, we plot $\tilde{F}$ as a
function of the ABCA$^\prime$ vesicle radius $R$ (the radius on
the outside of the vesicle at which the copolymer and solvent
densities are equal), which we vary by changing $V$ as detailed above. We fix the zero of our
free energy scale to correspond to a homogeneously-mixed system of the same
composition. The curve shows a minimum as a function
of $R$, demonstrating that the vesicles have a preferred
size. This is in sharp contrast to the monotonic decrease of the
free energy density of the
A$^\prime$CBA vesicle, plotted in the inset.

To understand this, note that, at smaller radii, the
free energy densities of both vesicles decrease with increasing size,
as the copolymers are less
compressed in the inner leaflet. The ABCA$^\prime$
vesicle always has a lower free energy than the A$^\prime$CBA vesicle. As the vesicle radius
increases, both membranes become flatter and the relative advantage of
the ABCA$^\prime$ vesicle decreases. The two lines then
approach each other, with the ABCA$^\prime$ curve now rising and the
A$^\prime$CBA curve continuing to fall. The monotonic form of
the A$^\prime$CBA curve is also clear evidence that the ABCA$^\prime$
minimum is not a finite size effect.

To calculate the relative polydispersity of the vesicles,
we plot the main free energy curve of Fig.\ \ref{min_fig} as a function
of $V$, and calculate
$\partial^2\tilde{F}/\partial V^2$ at the minimum. We calculate the
aggregate volume $v_{\text{m}}$ by integrating over the vesicle
density profile and subtracting the local volume fraction at the edge
of the system, where it has reached a stable bulk value. Next, we estimate the volume $v$ of a single
copolymer molecule. By recalling that all volumes are measured in
units of $a^3N^{3/2}$, and defining the segment volume such that
$v=a^3N$, we can show that $\Delta/p$ is given by the product of a term specified
uniquely by our SCFT calculations and $1/N^{1/4}$. This shows that the
polydispersity is rather insensitive to the choice of $N$ within the physical range \cite{gomez} of $N\sim
100-1000$. Even using the smallest value, $N=100$, we find clear
size selection, with $\Delta/p\approx 0.09$. Since the vesicle is
relatively flat, we can assume that its surface area is proportional
to $p$ and hence that $R\propto p^{1/2}$. This
yields a relative polydispersity of the radius of $0.05$, and
shows that strong size selection takes place in our simple model.

\begin{figure}
\includegraphics[width=2.9in]{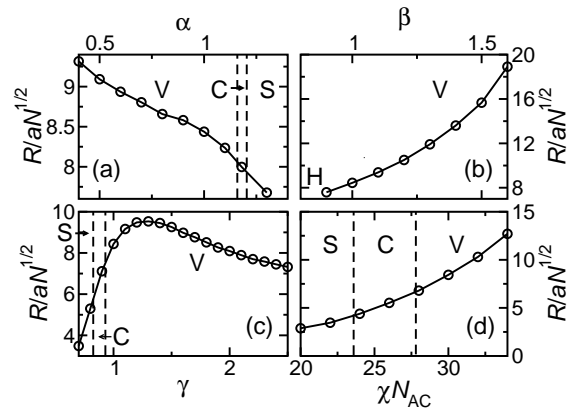}
\caption{\label{radii_fig} Vesicle radius as a function of the various
block sizes and $\chi_{\text{AC}}$. The radii $R$ are measured in units of $aN^{1/2}$, where
$N$ is the number of monomers in the original tetrablock copolymers
used in the calculations shown in Fig.\ \ref{profiles_fig} and Fig.\
\ref{min_fig}. The phase boundaries between the vesicle (V), cylinder
(C) and sphere (S) are calculated by interpolating
the free energy curves. In the region marked H, the copolymers are
homogeneously dispersed in the solvent.}
\end{figure}

We now investigate how the copolymer architecture controls the optimum
vesicle radius. In addition, the free energies of the optimum
spherical and cylindrical micelle structures are calculated, and the
shape transition boundaries marked on the graph (although only the
vesicle radius is shown). 
First, we multiply the
number of hydrophilic monomers in a tetrablock by a factor
$\alpha$ (Fig.\ \ref{radii_fig}a). The sizes of the A and A$^\prime$ blocks are
changed equally, and the numbers of the hydrophobic B and C monomers are
left unaltered from our earlier calculations. This procedure follows the
experiments of Brannan and Bates \cite{brannan_bates}, and, indeed, we
reproduce their result of a decrease in $R$ as
$\alpha$ is increased (see Fig.\ \ref{radii_fig}a). This is strong
evidence that the size selection mechanism at work in their
experiments is captured by our model. Furthermore, for $\alpha\approx 1.15$,
the free energy of the cylindrical micelle drops below that of the vesicle. This agrees with the experimental observation
\cite{brannan_bates} of the appearance of
cylindrical micelles as the A-block length is increased beyond a critical value.

Next, we multiply the number of B-monomers by a
factor $\beta$ (Fig.\ \ref{radii_fig}b), with the numbers of A and
C monomers kept constant at the values used in our earlier
calculations. For $\beta<0.9$, 
aggregates of all geometries considered become unstable. As $\beta$ is increased above this value,
the vesicle radius
grows rapidly, to reduce
compression of the B-blocks on the inner membrane surface. The relative
polydispersity in the radius falls as $\beta$ increases, reaching
$0.03$ at $\beta=1.6$. However, this is offset by the fast growth of
the vesicle radius itself.

A much more promising route to controlling vesicle size and
polydispersity is to vary the less strongly hydrophobic block C at constant numbers of A and B monomers (Fig.\ \ref{radii_fig}c). For
short C-block lengths ($\gamma<1$), the repulsion between the two hydrophobic
blocks is not sufficiently strong for the well-defined
structure shown in Fig.\ \ref{ditetra_fig} to form. Although vesicles
still exist as a solution to SCFT, they have higher free energies than
micelles and are small and strongly polydisperse. However, as $\gamma$ is increased above unity, the vesicle
radius reaches a maximum and then slowly decreases. This can be
understood by noting that, once the vesicle structure has been established, the C-blocks lie in the outer
half of the vesicle membrane. Increasing the length of the C-blocks
further then leads to the formation of smaller, more curved
aggregates, to reduce compression of the chains in this outer
layer. Furthermore, the relative polydispersity falls as $\gamma$ is
increased, remaining close to $0.03$ for $\gamma>1.4$. This result
shows that, once tetrablock vesicles have been formed in an experiment
\cite{brannan_bates}, the formation of small vesicles with a narrow
size distribution can be encouraged by lengthening the C-block. We
note that the physical mechanism behind this phenomenon is not
specific to our current model and can be expected to generalize to
systems with other solvents, such as water. An additional benefit of
this approach is that, due to the strong repulsion between the long
C-block and the other sections of the copolymer, the micellar
structures are strongly suppressed, disappearing altogether as
solutions to the SCFT at large $\gamma$.

Finally, in Fig.\ \ref{radii_fig}d, we show that increasing the
repulsion between A and C blocks from $\chi_{\text{AC}}=20/N$ to
$\chi_{\text{AC}}=34/N$ (which also increases $\chi_{\text{AB}}$) produces shape transitions between spherical
micelles, cylindrical micelles and vesicles. This demonstrates that
the repulsions between the various species must be above a certain
threshold for vesicles to form rather than micelles, where the blocks mix in
the corona. 

In summary, we have demonstrated that the curvature of membranes in
solution can be controlled by the architecture of the constituent
amphiphilic molecules. We use coarse-grained
calculations to show that copolymers composed of two central
hydrophobic blocks and two outer hydrophilic blocks form vesicles with
a preferred radius. To our knowledge, this is the only system where
the molecular structure of the amphiphiles can be shown directly to
fix the curvature of a membrane in solution. Control of curvature has
only been achieved before by mixing two types of amphiphile or by
using a complex self-assembly method. We have reproduced the
dependence of vesicle size on hydrophilic block length observed
experimentally \cite{brannan_bates}, and have shown how to encourage
the formation of vesicles with a narrow size distribution.

The most promising future direction is to focus on the
optimum region of parameter space identified here and to include more molecular detail
\cite{pogodin}, to tune the
polymer parameters to produce nearly monodisperse vesicles.


\begin{thebibliography}{33}
\expandafter\ifx\csname natexlab\endcsname\relax\def\natexlab#1{#1}\fi
\expandafter\ifx\csname bibnamefont\endcsname\relax
  \def\bibnamefont#1{#1}\fi
\expandafter\ifx\csname bibfnamefont\endcsname\relax
  \def\bibfnamefont#1{#1}\fi
\expandafter\ifx\csname citenamefont\endcsname\relax
  \def\citenamefont#1{#1}\fi
\expandafter\ifx\csname url\endcsname\relax
  \def\url#1{\texttt{#1}}\fi
\expandafter\ifx\csname urlprefix\endcsname\relax\def\urlprefix{URL }\fi
\providecommand{\bibinfo}[2]{#2}
\providecommand{\eprint}[2][]{\url{#2}}

\bibitem[{\citenamefont{He and Schmid}(2008)}]{he_schmid}
\bibinfo{author}{\bibfnamefont{X.}~\bibnamefont{He}} \bibnamefont{and}
  \bibinfo{author}{\bibfnamefont{F.}~\bibnamefont{Schmid}},
  \bibinfo{journal}{Phys.\ Rev.\ Lett.} \textbf{\bibinfo{volume}{100}},
  \bibinfo{pages}{137802} (\bibinfo{year}{2008}).

\bibitem[{\citenamefont{Jain and Bates}(2003)}]{jain_bates}
\bibinfo{author}{\bibfnamefont{S.}~\bibnamefont{Jain}} \bibnamefont{and}
  \bibinfo{author}{\bibfnamefont{F.~S.} \bibnamefont{Bates}},
  \bibinfo{journal}{Science} \textbf{\bibinfo{volume}{300}},
  \bibinfo{pages}{460} (\bibinfo{year}{2003}).

\bibitem[{\citenamefont{Zhang and Eisenberg}(1996)}]{zhang_eisenberg}
\bibinfo{author}{\bibfnamefont{L.~F.} \bibnamefont{Zhang}} \bibnamefont{and}
  \bibinfo{author}{\bibfnamefont{A.}~\bibnamefont{Eisenberg}},
  \bibinfo{journal}{Science} \textbf{\bibinfo{volume}{272}},
  \bibinfo{pages}{1777} (\bibinfo{year}{1996}).

\bibitem[{\citenamefont{Haleva and Diamant}(2008)}]{haleva_diamant}
\bibinfo{author}{\bibfnamefont{E.}~\bibnamefont{Haleva}} \bibnamefont{and}
  \bibinfo{author}{\bibfnamefont{H.}~\bibnamefont{Diamant}},
  \bibinfo{journal}{Phys.\ Rev.\ Lett.} \textbf{\bibinfo{volume}{101}},
  \bibinfo{pages}{078104} (\bibinfo{year}{2008}).

\bibitem[{\citenamefont{Choi et~al.}(2010)\citenamefont{Choi, Lodge, and
  Bates}}]{choi}
\bibinfo{author}{\bibfnamefont{S.-H.} \bibnamefont{Choi}},
  \bibinfo{author}{\bibfnamefont{T.~P.} \bibnamefont{Lodge}}, \bibnamefont{and}
  \bibinfo{author}{\bibfnamefont{F.~S.} \bibnamefont{Bates}},
  \bibinfo{journal}{Phys.\ Rev.\ Lett.} \textbf{\bibinfo{volume}{104}},
  \bibinfo{pages}{047802} (\bibinfo{year}{2010}).

\bibitem[{\citenamefont{Lee et~al.}(2006)\citenamefont{Lee, Agarwal, Bose,
  Payne, and Raghavan}}]{lee_agarwal}
\bibinfo{author}{\bibfnamefont{J.~H.} \bibnamefont{Lee}},
  \bibinfo{author}{\bibfnamefont{V.}~\bibnamefont{Agarwal}},
  \bibinfo{author}{\bibfnamefont{A.}~\bibnamefont{Bose}},
  \bibinfo{author}{\bibfnamefont{G.~F.} \bibnamefont{Payne}}, \bibnamefont{and}
  \bibinfo{author}{\bibfnamefont{S.~R.} \bibnamefont{Raghavan}},
  \bibinfo{journal}{Phys.\ Rev.\ Lett.} \textbf{\bibinfo{volume}{96}},
  \bibinfo{pages}{048102} (\bibinfo{year}{2006}).

\bibitem[{\citenamefont{Kaler et~al.}(1989)\citenamefont{Kaler, Murthy,
  Rodriguez, and Zasadzinski}}]{kaler}
\bibinfo{author}{\bibfnamefont{E.~W.} \bibnamefont{Kaler}},
  \bibinfo{author}{\bibfnamefont{A.~K.} \bibnamefont{Murthy}},
  \bibinfo{author}{\bibfnamefont{B.~E.} \bibnamefont{Rodriguez}},
  \bibnamefont{and} \bibinfo{author}{\bibfnamefont{J.~A.~N.}
  \bibnamefont{Zasadzinski}}, \bibinfo{journal}{Science}
  \textbf{\bibinfo{volume}{245}}, \bibinfo{pages}{1371} (\bibinfo{year}{1989}).

\bibitem[{\citenamefont{Katagiri and Caruso}(2005)}]{katagiri}
\bibinfo{author}{\bibfnamefont{K.}~\bibnamefont{Katagiri}} \bibnamefont{and}
  \bibinfo{author}{\bibfnamefont{F.}~\bibnamefont{Caruso}},
  \bibinfo{journal}{Adv.\ Mater.} \textbf{\bibinfo{volume}{17}},
  \bibinfo{pages}{738} (\bibinfo{year}{2005}).

\bibitem[{\citenamefont{Li et~al.}(2009)\citenamefont{Li, Pr\'{e}vost,
  Schweins, Marcelis, Leermakers, Stuart, and Sudh\"{o}lter}}]{li_cohen-stuart}
\bibinfo{author}{\bibfnamefont{F.}~\bibnamefont{Li}},
  \bibinfo{author}{\bibfnamefont{S.}~\bibnamefont{Pr\'{e}vost}},
  \bibinfo{author}{\bibfnamefont{R.}~\bibnamefont{Schweins}},
  \bibinfo{author}{\bibfnamefont{A.~T.~M.} \bibnamefont{Marcelis}},
  \bibinfo{author}{\bibfnamefont{F.~A.~M.} \bibnamefont{Leermakers}},
  \bibinfo{author}{\bibfnamefont{M.~A.~C.} \bibnamefont{Stuart}},
  \bibnamefont{and} \bibinfo{author}{\bibfnamefont{E.~J.~R.}
  \bibnamefont{Sudh\"{o}lter}}, \bibinfo{journal}{Soft Matter}
  \textbf{\bibinfo{volume}{5}}, \bibinfo{pages}{4169} (\bibinfo{year}{2009}).

\bibitem[{\citenamefont{Nieh et~al.}(2003)\citenamefont{Nieh, Harroun,
  Raghunathan, Glinka, and Katsaras}}]{nieh}
\bibinfo{author}{\bibfnamefont{M.-P.} \bibnamefont{Nieh}},
  \bibinfo{author}{\bibfnamefont{T.~A.} \bibnamefont{Harroun}},
  \bibinfo{author}{\bibfnamefont{V.~A.} \bibnamefont{Raghunathan}},
  \bibinfo{author}{\bibfnamefont{C.~J.} \bibnamefont{Glinka}},
  \bibnamefont{and} \bibinfo{author}{\bibfnamefont{J.}~\bibnamefont{Katsaras}},
  \bibinfo{journal}{Phys.\ Rev.\ Lett.} \textbf{\bibinfo{volume}{91}},
  \bibinfo{pages}{158105} (\bibinfo{year}{2003}).

\bibitem[{\citenamefont{Joannic et~al.}(1997)\citenamefont{Joannic, Auvray, and
  Lasic}}]{joannic}
\bibinfo{author}{\bibfnamefont{R.}~\bibnamefont{Joannic}},
  \bibinfo{author}{\bibfnamefont{L.}~\bibnamefont{Auvray}}, \bibnamefont{and}
  \bibinfo{author}{\bibfnamefont{D.~D.} \bibnamefont{Lasic}},
  \bibinfo{journal}{Phys.\ Rev.\ Lett.} \textbf{\bibinfo{volume}{78}},
  \bibinfo{pages}{3402} (\bibinfo{year}{1997}).

\bibitem[{\citenamefont{Campelo and Hernandez-Machado}(2008)}]{campelo}
\bibinfo{author}{\bibfnamefont{F.}~\bibnamefont{Campelo}} \bibnamefont{and}
  \bibinfo{author}{\bibfnamefont{A.}~\bibnamefont{Hernandez-Machado}},
  \bibinfo{journal}{Phys.\ Rev.\ Lett.} \textbf{\bibinfo{volume}{100}},
  \bibinfo{pages}{158103} (\bibinfo{year}{2008}).

\bibitem[{\citenamefont{Safran et~al.}(1990)\citenamefont{Safran, Pincus, and
  Andelman}}]{safran_pincus_andelman}
\bibinfo{author}{\bibfnamefont{S.~A.} \bibnamefont{Safran}},
  \bibinfo{author}{\bibfnamefont{P.}~\bibnamefont{Pincus}}, \bibnamefont{and}
  \bibinfo{author}{\bibfnamefont{D.}~\bibnamefont{Andelman}},
  \bibinfo{journal}{Science} \textbf{\bibinfo{volume}{248}},
  \bibinfo{pages}{354} (\bibinfo{year}{1990}).

\bibitem[{\citenamefont{Howse et~al.}(2009)\citenamefont{Howse, Jones,
  Battaglia, Ducker, Leggett, and Ryan}}]{howse}
\bibinfo{author}{\bibfnamefont{J.~R.} \bibnamefont{Howse}},
  \bibinfo{author}{\bibfnamefont{R.~A.~L.} \bibnamefont{Jones}},
  \bibinfo{author}{\bibfnamefont{G.}~\bibnamefont{Battaglia}},
  \bibinfo{author}{\bibfnamefont{R.~E.} \bibnamefont{Ducker}},
  \bibinfo{author}{\bibfnamefont{G.~J.} \bibnamefont{Leggett}},
  \bibnamefont{and} \bibinfo{author}{\bibfnamefont{A.~J.} \bibnamefont{Ryan}},
  \bibinfo{journal}{Nat.\ Mater.} \textbf{\bibinfo{volume}{8}},
  \bibinfo{pages}{507} (\bibinfo{year}{2009}).

\bibitem[{\citenamefont{Rank et~al.}(2009)\citenamefont{Rank, Hauschild,
  F\"{o}rster, and Schubert}}]{rank}
\bibinfo{author}{\bibfnamefont{A.}~\bibnamefont{Rank}},
  \bibinfo{author}{\bibfnamefont{S.}~\bibnamefont{Hauschild}},
  \bibinfo{author}{\bibfnamefont{S.}~\bibnamefont{F\"{o}rster}},
  \bibnamefont{and} \bibinfo{author}{\bibfnamefont{R.}~\bibnamefont{Schubert}},
  \bibinfo{journal}{Langmuir} \textbf{\bibinfo{volume}{25}},
  \bibinfo{pages}{1337} (\bibinfo{year}{2009}).

\bibitem[{\citenamefont{Taylor et~al.}(2003)\citenamefont{Taylor, Xu, Fletcher,
  and Paunov}}]{taylor}
\bibinfo{author}{\bibfnamefont{P.}~\bibnamefont{Taylor}},
  \bibinfo{author}{\bibfnamefont{C.}~\bibnamefont{Xu}},
  \bibinfo{author}{\bibfnamefont{P.~D.~I.} \bibnamefont{Fletcher}},
  \bibnamefont{and} \bibinfo{author}{\bibfnamefont{V.~N.}
  \bibnamefont{Paunov}}, \bibinfo{journal}{Chem. Commun.} pp.
  \bibinfo{pages}{1732--1733} (\bibinfo{year}{2003}).

\bibitem[{\citenamefont{Thiele et~al.}(2010)\citenamefont{Thiele, Steinhauser,
  Pfohl, and F\"{o}rster}}]{thiele}
\bibinfo{author}{\bibfnamefont{J.}~\bibnamefont{Thiele}},
  \bibinfo{author}{\bibfnamefont{D.}~\bibnamefont{Steinhauser}},
  \bibinfo{author}{\bibfnamefont{T.}~\bibnamefont{Pfohl}}, \bibnamefont{and}
  \bibinfo{author}{\bibfnamefont{S.}~\bibnamefont{F\"{o}rster}},
  \bibinfo{journal}{Langmuir} \textbf{\bibinfo{volume}{26}},
  \bibinfo{pages}{6860} (\bibinfo{year}{2010}).

\bibitem[{\citenamefont{Zhu and Szostak}(2009)}]{zhu}
\bibinfo{author}{\bibfnamefont{T.~F.} \bibnamefont{Zhu}} \bibnamefont{and}
  \bibinfo{author}{\bibfnamefont{J.~W.} \bibnamefont{Szostak}},
  \bibinfo{journal}{PLoS ONE} \textbf{\bibinfo{volume}{4}},
  \bibinfo{pages}{e5009} (\bibinfo{year}{2009}).

\bibitem[{\citenamefont{Brannan and Bates}(2004)}]{brannan_bates}
\bibinfo{author}{\bibfnamefont{A.~K.} \bibnamefont{Brannan}} \bibnamefont{and}
  \bibinfo{author}{\bibfnamefont{F.~S.} \bibnamefont{Bates}},
  \bibinfo{journal}{Macromolecules} \textbf{\bibinfo{volume}{37}},
  \bibinfo{pages}{8816} (\bibinfo{year}{2004}).

\bibitem[{\citenamefont{Gomez et~al.}(2005)\citenamefont{Gomez, Rappl, Agarwal,
  Bose, Schmutz, Marques, and Balsara}}]{gomez}
\bibinfo{author}{\bibfnamefont{E.~D.} \bibnamefont{Gomez}},
  \bibinfo{author}{\bibfnamefont{T.~J.} \bibnamefont{Rappl}},
  \bibinfo{author}{\bibfnamefont{V.}~\bibnamefont{Agarwal}},
  \bibinfo{author}{\bibfnamefont{A.}~\bibnamefont{Bose}},
  \bibinfo{author}{\bibfnamefont{M.}~\bibnamefont{Schmutz}},
  \bibinfo{author}{\bibfnamefont{C.~M.} \bibnamefont{Marques}},
  \bibnamefont{and} \bibinfo{author}{\bibfnamefont{N.~P.}
  \bibnamefont{Balsara}}, \bibinfo{journal}{Macromolecules}
  \textbf{\bibinfo{volume}{38}}, \bibinfo{pages}{3567} (\bibinfo{year}{2005}).

\bibitem[{\citenamefont{Cui and Jiang}(2011)}]{cui}
\bibinfo{author}{\bibfnamefont{J.}~\bibnamefont{Cui}} \bibnamefont{and}
  \bibinfo{author}{\bibfnamefont{W.}~\bibnamefont{Jiang}},
  \bibinfo{journal}{Langmuir} \textbf{\bibinfo{volume}{27}},
  \bibinfo{pages}{10141} (\bibinfo{year}{2011}).

\bibitem[{\citenamefont{Kinning et~al.}(1988)\citenamefont{Kinning, Winey, and
  Thomas}}]{kinning_winey_thomas}
\bibinfo{author}{\bibfnamefont{D.~J.} \bibnamefont{Kinning}},
  \bibinfo{author}{\bibfnamefont{K.~I.} \bibnamefont{Winey}}, \bibnamefont{and}
  \bibinfo{author}{\bibfnamefont{E.~L.} \bibnamefont{Thomas}},
  \bibinfo{journal}{Macromolecules} \textbf{\bibinfo{volume}{21}},
  \bibinfo{pages}{3502} (\bibinfo{year}{1988}).

\bibitem[{\citenamefont{Greenall et~al.}(2009)\citenamefont{Greenall, Buzza,
  and McLeish}}]{gbm_jcp}
\bibinfo{author}{\bibfnamefont{M.~J.} \bibnamefont{Greenall}},
  \bibinfo{author}{\bibfnamefont{D.~M.~A.} \bibnamefont{Buzza}},
  \bibnamefont{and} \bibinfo{author}{\bibfnamefont{T.~C.~B.}
  \bibnamefont{McLeish}}, \bibinfo{journal}{J. Chem.\ Phys.}
  \textbf{\bibinfo{volume}{131}}, \bibinfo{pages}{034904}
  (\bibinfo{year}{2009}).

\bibitem[{\citenamefont{Katsov et~al.}(2004)\citenamefont{Katsov, M\"{u}ller,
  and Schick}}]{katsov1}
\bibinfo{author}{\bibfnamefont{K.}~\bibnamefont{Katsov}},
  \bibinfo{author}{\bibfnamefont{M.}~\bibnamefont{M\"{u}ller}},
  \bibnamefont{and} \bibinfo{author}{\bibfnamefont{M.}~\bibnamefont{Schick}},
  \bibinfo{journal}{Biophys.\ J.} \textbf{\bibinfo{volume}{87}},
  \bibinfo{pages}{3277} (\bibinfo{year}{2004}).

\bibitem[{\citenamefont{Papanicolaou et~al.}(2012)\citenamefont{Papanicolaou,
  Phillippo, and Walsh}}]{papanicolaou}
\bibinfo{author}{\bibfnamefont{K.~N.} \bibnamefont{Papanicolaou}},
  \bibinfo{author}{\bibfnamefont{M.~M.} \bibnamefont{Phillippo}},
  \bibnamefont{and} \bibinfo{author}{\bibfnamefont{K.}~\bibnamefont{Walsh}},
  \bibinfo{journal}{Am.\ J. Physio.\ Heart Circ.\ Physiol.}
  \textbf{\bibinfo{volume}{303}}, \bibinfo{pages}{H243} (\bibinfo{year}{2012}).

\bibitem[{\citenamefont{Schmid}(2011)}]{boudenne_book}
\bibinfo{author}{\bibfnamefont{F.}~\bibnamefont{Schmid}}, in
  \emph{\bibinfo{booktitle}{Handbook of Multiphase Polymer Systems}}, edited by
  \bibinfo{editor}{\bibfnamefont{A.}~\bibnamefont{Boudenne}}
  (\bibinfo{publisher}{John Wiley and Sons}, \bibinfo{address}{Chichester},
  \bibinfo{year}{2011}), chap.~\bibinfo{chapter}{3}.

\bibitem[{\citenamefont{Edwards}(1965)}]{edwards}
\bibinfo{author}{\bibfnamefont{S.~F.} \bibnamefont{Edwards}},
  \bibinfo{journal}{Proc.\ Phys.\ Soc.} \textbf{\bibinfo{volume}{85}},
  \bibinfo{pages}{613} (\bibinfo{year}{1965}).

\bibitem[{\citenamefont{Matsen}(2006)}]{matsen_book}
\bibinfo{author}{\bibfnamefont{M.~W.} \bibnamefont{Matsen}}, in
  \emph{\bibinfo{booktitle}{Soft Matter}}, edited by
  \bibinfo{editor}{\bibfnamefont{G.}~\bibnamefont{Gompper}} \bibnamefont{and}
  \bibinfo{editor}{\bibfnamefont{M.}~\bibnamefont{Schick}}
  (\bibinfo{publisher}{Wiley-VCH}, \bibinfo{address}{Weinheim},
  \bibinfo{year}{2006}), chap.~\bibinfo{chapter}{2}.

\bibitem[{\citenamefont{Cavallo et~al.}(2006)\citenamefont{Cavallo, M\"{u}ller,
  and Binder}}]{cavallo}
\bibinfo{author}{\bibfnamefont{A.}~\bibnamefont{Cavallo}},
  \bibinfo{author}{\bibfnamefont{M.}~\bibnamefont{M\"{u}ller}},
  \bibnamefont{and} \bibinfo{author}{\bibfnamefont{K.}~\bibnamefont{Binder}},
  \bibinfo{journal}{Macromolecules} \textbf{\bibinfo{volume}{39}},
  \bibinfo{pages}{9539} (\bibinfo{year}{2006}).

\bibitem[{\citenamefont{Kim and Matsen}(2009)}]{kim_matsen}
\bibinfo{author}{\bibfnamefont{J.~U.} \bibnamefont{Kim}} \bibnamefont{and}
  \bibinfo{author}{\bibfnamefont{M.~W.} \bibnamefont{Matsen}},
  \bibinfo{journal}{Phys. Rev.\ Lett.} \textbf{\bibinfo{volume}{102}},
  \bibinfo{pages}{078303} (\bibinfo{year}{2009}).

\bibitem[{\citenamefont{Puvvada and Blankschtein}(1990)}]{puvvada}
\bibinfo{author}{\bibfnamefont{S.}~\bibnamefont{Puvvada}} \bibnamefont{and}
  \bibinfo{author}{\bibfnamefont{D.}~\bibnamefont{Blankschtein}},
  \bibinfo{journal}{J. Chem.\ Phys.} \textbf{\bibinfo{volume}{92}},
  \bibinfo{pages}{3710} (\bibinfo{year}{1990}).

\bibitem[{\citenamefont{Ulrich et~al.}(2009)\citenamefont{Ulrich, Galvosas,
  K\"{a}rger, and Grinberg}}]{ulrich}
\bibinfo{author}{\bibfnamefont{K.}~\bibnamefont{Ulrich}},
  \bibinfo{author}{\bibfnamefont{P.}~\bibnamefont{Galvosas}},
  \bibinfo{author}{\bibfnamefont{J.}~\bibnamefont{K\"{a}rger}},
  \bibnamefont{and} \bibinfo{author}{\bibfnamefont{F.}~\bibnamefont{Grinberg}},
  \bibinfo{journal}{Phys.\ Rev.\ Lett.} \textbf{\bibinfo{volume}{102}},
  \bibinfo{pages}{037801} (\bibinfo{year}{2009}).

\bibitem[{\citenamefont{Pogodin and Baulin}(2010)}]{pogodin}
\bibinfo{author}{\bibfnamefont{S.}~\bibnamefont{Pogodin}} \bibnamefont{and}
  \bibinfo{author}{\bibfnamefont{V.~A.} \bibnamefont{Baulin}},
  \bibinfo{journal}{Soft Matter} \textbf{\bibinfo{volume}{6}},
  \bibinfo{pages}{2216} (\bibinfo{year}{2010}).

\end{thebibliography}
\end{document}